\documentclass[12pt]{article}

\usepackage{amsmath,amsfonts,epsfig}
\usepackage{amsmath,amssymb,amsthm,amscd}
\input epsf.sty

\newcommand{\Tr}{\textrm{Tr}}

\usepackage{amssymb,graphicx}

\usepackage{epsf}
\usepackage{graphicx,epsfig}
\usepackage{amsfonts}
\usepackage{amssymb}
\usepackage{cite}

\def\etmiss{E\!\!\!\!\slash_{T}}

\makeatletter
\renewcommand\section{\@startsection {section}{1}{\z@}%
                                 {-3.5ex \@plus -1ex \@minus -.2ex}%nn
                                   {2.3ex \@plus.2ex}%
                                   {\normalfont\large\bfseries}}
\renewcommand\subsection{\@startsection{subsection}{2}{\z@}%
                                   {-3.25ex\@plus -1ex \@minus -.2ex}%
                                     {1.5ex \@plus .2ex}%
                                     {\normalfont\bfseries}}
\renewcommand\subsubsection{\@startsection{subsubsection}{3}{\z@}%
                                   {-3.25ex\@plus -1ex \@minus -.2ex}%
                                     {1.5ex \@plus .2ex}%
                                     {\normalfont\itshape}}
\makeatother

\newcommand{\Letter}{
\setlength{\textwidth}{16.5cm}
   \setlength{\textheight}{22.6cm}
    \hoffset=-0.6in
\voffset=-2.1cm }

\Letter

\begin{document}
\newcommand{\be}{\begin{equation}}
\newcommand{\ee}{\end{equation}}
\newcommand{\bea}{\begin{eqnarray}}
\newcommand{\eea}{\end{eqnarray}}
\newcommand{\barr}{\begin{array}}
\newcommand{\earr}{\end{array}}

\thispagestyle{empty}
\begin{flushright}
\parbox[t]{1.5in}{
MADPH-10-1557 \\
MAD-TH-09-03\\
IPMU10-0043}
\end{flushright}

\vspace*{0.5in}

\begin{center}
{\Large \bf Top Quarks as a Window to String Resonances}

\vspace*{0.5in} {\large Zhe Dong$^{1}$, Tao Han$^{1}$, Min-xin Huang$^2$, and Gary Shiu$^{1,3}$}
\\[.3in]
{\em     ${}^1$ Department of Physics,
     University of Wisconsin,
     Madison, WI 53706, USA\\[0.1in]
${}^2$ Institute for the Physics and Mathematics of the Universe (IPMU),  \\ University of Tokyo, Kashiwa, Chiba 277-8582, Japan \\[0.1in]
${}^3$ Institute for Advanced Study, Hong Kong University of Science and Technology\\
Hong Kong, People's Republic of China} 
\\[0.3in]
\end{center}

\begin{center}
{\bf
Abstract}
\end{center}
\noindent
We study the discovery potential of string resonances decaying to $t\bar t$ final state at the LHC.
We point out that top quark pair production is a promising and an advantageous channel for studying such resonances, due to their low Standard Model background and unique kinematics. 
We study the invariant mass distribution and angular dependence of 
the top pair production cross section via exchanges of string resonances.
The mass ratios of these resonances and the unusual angular distribution may help identify their fundamental properties
and distinguish them from other new physics. We find that string resonances for a string scale below 4 TeV can be detected via the $t\bar t$ channel, either from reconstructing the $t \overline{t}$ semi-leptonic decay or recent techniques in identifying highly boosted tops.

\vfill

\newpage

\tableofcontents

\section{Introduction}

With the turn-on of the CERN Large Hadron Collider (LHC), a new era of discovery has just begun. 
This is an opportune time to explore and anticipate
various exotic signatures of physics beyond the Standard Model (BSM) that could potentially be revealed at
the LHC.  
Arguably, a major driving force behind the consideration of BSM physics is the hierarchy problem,
{\it i.e.,}  the puzzle of why there is such a
 huge disparity between the electroweak scale and the apparent scale of quantum gravity.
Thus, it  is of  interest to identify what kind of novel signatures  can plausibly arise in
theories  with the  capacity to describe physics over this enormous range of scales.
String theory being our most developed theory of quantum gravity 
provides a perfect arena for such investigations.

Despite this promise, attempts to extract collider signatures of string theory
have been plagued with difficulties. First of all, the energy scale associated with string theory, known as the string scale $M_s \sim 1/\sqrt{\alpha'}$, is often assumed to be close to the Planck scale or the Grand Unification (GUT) scale  \cite{review-high-string-scale}, making {\it direct} signals difficult to access.
 Secondly, most if not all string constructions come equipped with  
 additional light
fields even in the point particle (i.e., $\alpha'   \rightarrow 0$) limit.
 These light BSM particles may include new gauge bosons, an extended Higgs sector, or matter with non-Standard Model charges. 
Their quantum numbers and couplings 
vary from model to model, and thus
their presence makes it difficult to separate the ``forest" (genuine stringy effects) from the ``trees"
(peculiarities of specific models).
Without reference to specific models, most studies 
resort to finding ``footprints" of string theory
on low energy observables
and the relations among them,
e.g.,
the pattern of the resulting soft supersymmetry Lagrangian \cite{footprints,Aparicio:2008wh,Conlon:2007xv}.

In recent years, however, it has become evident that the energy scale associated with string effects
can be significantly lower than the Planck or GUT scale.
With the advent of branes and fluxes (field strengths of generalized gauge fields), it is possible for
the extra dimensions in string theory
to be large
\cite{ADD,ShiuTye} (realizing concretely earlier suggestions \cite{Antoniadis:1990ew,Lykken}) or 
warped \cite{RS,Verlinde,Dasgupta,Greene,GKP,KKLT} while maintaining the observed gauge and gravitational couplings.
In the presence of large extra
dimensions, the fundamental string scale is lowered by the dilution of gravity.
For warped extra dimensions, the fundamental string scale remains high but there can be strongly warped regions in the internal space 
 where the {\it local} string scale is warped down. In either case, the energy scale associated with string effects is
greatly reduced and can in principle be within the reach of the current and upcoming collider experiments. 
Indeed, preliminary studies have demonstrated that 
if the string scale (fundamental or local)  is sufficiently low, such
string states can induce observable effects at the 
LHC \cite{Shiu:1999iw,Cullen:2000ef, BFH,Chialva:2005gt,LST,AGLNST,AGLNST2,Lust:2009pz,AGNT1,AGNT2,AGT3,Meade:2007sz,Hassanain:2009at}.

Another development which motivates our current studies is the interesting observation that
for a large class of string models where the Standard Model is realized on the worldvolume of 
D-branes\footnote{For some recent reviews on D-brane model building, see, e.g., \cite{Blumenhagen:2005mu,Blumenhagen:2006ci,Marchesano:2007de,Malyshev:2007zz}.},
the leading contributions to certain processes at hadron colliders are 
{\it universal} \cite{LST}.
This is because the string amplitudes which describe $2 \rightarrow 2$ parton scattering subprocesses involving 
four gluons as well as two gluons plus two quarks
are, to leading order in string coupling (but all orders in $\alpha'$),
 independent of the details of the compactification. More specifically,
the string corrections to such parton subprocesses are the same regardless of
 the configuration of branes,  the geometry of  the 
 extra dimensions, and whether supersymmetry is broken or not.
 This model-independence makes it possible to compute the leading string corrections to   dijet signals at the LHC 
 \cite{AGLNST}. 
 Naively, the four fermion subprocesses like quark-antiquark scattering, include (even in leading order
of the string coupling) also the exchanges of heavy Kaluza-Klein (KK) and winding states and hence they are model specific. However, their contribution to the dijet production computed in \cite{AGLNST} is color and parton distribution function (PDF) suppressed.
 Moreover, the $s$-channel excitation of string resonances is absent in these four fermion 
 amplitudes. Therefore, not only do these  four fermion subprocesses 
 not affect the universality of the
cross section around string resonances, the effective four-fermion contact terms generated by the KK recurrences can be used as discriminators of different string compactifications \cite{AGLNST2}. 
Furthermore, due to the structure of the Veneziano amplitude,
the effective four-fermion interactions resulting from integrating out the heavy string modes
come as  dimension-$8$ rather than the usual dimension-$6$ operators (and thus further suppressed by $s/M_s^2$), 
leading to a much weaker constraint on $M_s$ from precision electroweak tests \cite{Burikham:2003ha}.
Thus, within this general framework of D-brane models, one can cleanly
 extract the leading model-independent
genuine stringy effects,
while precision experiments may allow us to constrain the subleading model-dependent corrections
and hence the underlying string compactification.

In this paper, we revisit the prospects of detecting 
string resonances at the LHC in light of the above developments.
By considering various possible decay products 
of the string resonances, we identify a unique detection channel as $t \overline{t}$ 
production.
Besides 
the enhancement of 
quark production in comparison to the electroweak processes like diphoton or $ZZ$ by the
group (so called Chan-Paton) factors,
the Standard Model background for $t \overline{t}$ production is about $10^{-2}$ of a generic jet and thus provides a good signal to background ratio for detection of new physics (see e.g. \cite{Han:2008xb})).
In contrast to other quarks, top quarks promptly decay via weak interaction before QCD sets in for hadronization \cite{Bigi:1986jk}.
Rather than complicated bound states, the properties of ``bare" top quarks  may be accessible for scrutiny, e.g., through their semi-leptonic decay \cite{Han:2008xb}, or the more recently discussed methods to identify highly boosted tops.
Among the most distinctive features of string theory is the existence of a tower of excited states with increasing mass and spin, the so called Regge behavior.
The exchanges of such higher spin states lead to unusual angular distributions of various cross sections,
which can be used to distinguish these string resonances from other new BSM
massive particles such as $Z'$.
We investigate the discovery potential of string resonances at the LHC, with particular emphasis on 
$t \overline{t}$ production and their angular distributions.

This paper is organized as follows. In Section \ref{Amplitudes}, we discuss string theory amplitudes 
relevant for $t \overline{t}$ production at the LHC.  We decompose the cross sections
  in terms of the Wigner d-functions to facilitate the analysis of their angular distributions. 
 We also derive the decay widths of the first and second excited string resonances into Standard Model particles. In Section \ref{Results}, we present the results of our detailed phenomenological study for the signal final state of $t\bar t$
at the LHC. 
We further extend the signal study to include the $t\bar t g$ channel in Section 4, which leads to the possibility to discover 
both $n=1$ and 2 string states in the same event sample. We comment on the signal treatment for  a significantly heavier string state in Section 5.
We conclude in Section \ref{Conclusion}.

\section{$t\bar{t}$ Production via a String Resonance}\label{Amplitudes}

Let us start with the string amplitudes relevant to $t \overline{t}$ production.
At the parton level,
the main contribution 
comes from
the $gg\rightarrow t\bar{t}$ subprocess since, as explained in \cite{AGLNST}, the $q\bar{q}\rightarrow t\bar{t}$ amplitude is suppressed compared to gluon fusion by both color group factors and 
 parton luminosity.  We can adopt the gluon fusion amplitude computed in \cite{AGLNST} as it 
 does not distinguish, for the purpose of our discussions\footnote{As we will discuss in Section \ref{Conclusion}, model-dependent processes such as four-fermion amplitudes can distinguish different quarks as well as their chiralities.}, different types of quarks:
\begin{eqnarray}
 && |\mathcal{M}(gg\rightarrow t\bar{t})|^2 = \frac{g^4}{6}\frac{t^2+u^2}{s^2} \left[\frac{1}{ut} (tV_t+uV_u)^2 - \frac{9}{4}  V_tV_u \right], \\
 && 
 V_t \equiv V(s,t,u)\equiv \Gamma(1-\frac{s}{M_s^2})\Gamma(1-\frac{u}{M_s^2})/\Gamma(1+\frac{t}{M_s^2}),\quad  
 V_u \equiv V(s,u,t) ,
 \end{eqnarray}  
 and for completeness $V_s \equiv V (t,s,u)$. The Veneziano amplitude $V$ may develop simple poles near the
 Regge resonances. In the following, we analyze this amplitude around the $n=1,2$ resonances.
 
 \subsection{$n=1$ Resonances}

There are no massless particles propagating in the $s$-channel in the energy regime far below the string scale. We will focus on the first Regge string resonance when $s\rightarrow M_s^2$. 
Expanding the expression around $s=M_s^2$, we have for $n=1$
 \begin{eqnarray}
|\mathcal{M}(gg\rightarrow t\bar{t})|^2 = \frac{7}{24}\frac{g^4}{M_s^4} \left[ 0.24 \frac{ut(u^2+t^2)}{(s-M_s^2)^2+(\Gamma_{g^*}^{J=2}M_s)^2}
+0.76 \frac{ut(u^2+t^2)}{(s-M_s^2)^2+(\Gamma_{C^*}^{J=2}M_s)^2} \right]
\nonumber \\
\end{eqnarray}
where $g^*$ and $C^*$ label the string resonances of $SU(3)$ gluon and $U(1)$ gauge boson on the $U(3)$ QCD bane stack. We have included their decay widths to regularize the resonances. The decay rates are given by \cite{AGT3}
\begin{equation}
\Gamma_{g^*}^{J=2}=45\ ({M_s \over {\rm 1\ TeV} })\ {\rm GeV },\quad {\rm and}\ 
\Gamma_{C^*}^{J=2}=75\ ({M_s\over {\rm 1\ TeV}})\ {\rm GeV}. 
\end{equation}

\begin{figure}[tb]
\begin{center}
\epsfysize=2.1in\epsfbox{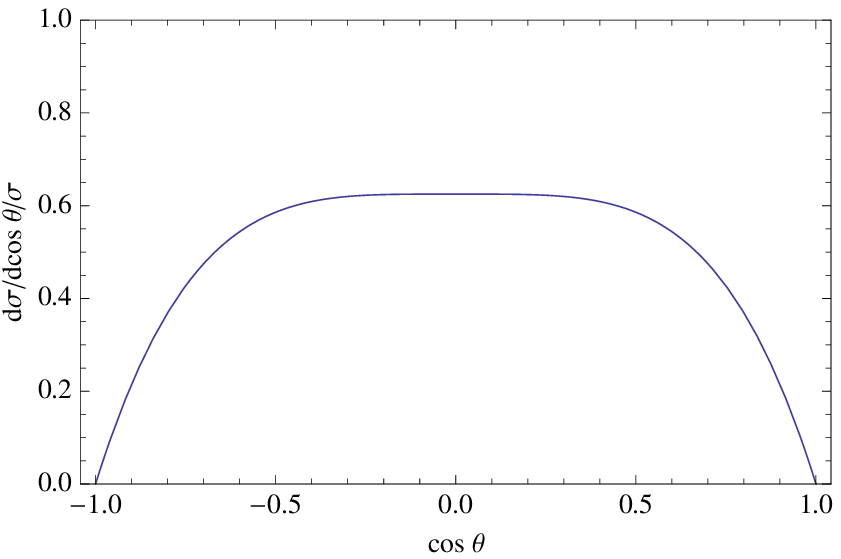}
\epsfysize=2.05in\epsfbox{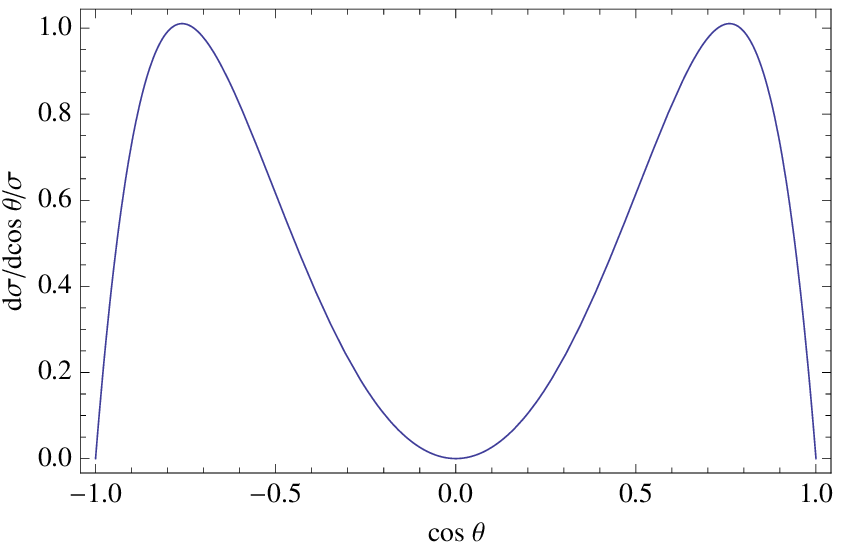}
\caption{The normalized angular distribution of $gg \rightarrow t \bar{t}$ via the exchange of a string resonance
for (a) $n=1$ (left), and (b) $n=2$ (right).} 
\label{figure4}
\end{center}
\end{figure}

Let the angle between the outgoing $t$ quark  and the scattering axis ($\hat z$) be $\theta$, the Mandelstam variables $t$ and $u$ can be written as
\begin{eqnarray} \label{tu}
t=-\frac{s}{2}(1+\cos \theta ),~~~ u=-\frac{s}{2}(1-\cos \theta ) .
\end{eqnarray}
It is intuitive to see the feature of the amplitudes in terms of the contributions from
states with a fixed angular momentum. 
The amplitudes can be decomposed as follows 
\begin{eqnarray}
{\cal M} (s,t,u)= 16 \pi \sum_{j,m,m^\prime}(2j+1) a_{j,m,m^\prime} (s) d^j_{m,m^\prime}(\theta) ,
\end{eqnarray}
where the Wigner $d$-function $d^j_{m,m^\prime}(\theta)$ signifies a state of total angular momentum $j$,
with $m\ (m')$ the helicity difference of the initial (final) state particles. 
We thus find 
\begin{eqnarray}
 |\mathcal{M}(gg\rightarrow t\bar{t})|^2 = \frac{7}{96} g^4 M_s^4 \left[ 0.24 ~\frac{|d^2_{2,1}(\theta)|^2+|d^2_{2,-1}(\theta)|^2}{(s-M_s^2)^2+(\Gamma_{g^*}^{J=2}M_s)^2}  +0.76 ~\frac{|d^2_{2,1}(\theta)|^2+|d^2_{2,-1}(\theta)|^2}{(s-M_s^2)^2+(\Gamma_{C^*}^{J=2}M_s)^2} \right] .
 \nonumber \\
\end{eqnarray}
It yields a spin $J=2$  resonance, with a dominant P-wave behavior. 
The angular distribution of  the $g g \rightarrow t \bar{t}$ amplitude for $n=1$ is 
depicted in Fig.~\ref{figure4}(a).

The integrated cross section is, for $m_t \ll M_{s}$, 
\begin{eqnarray}
\sigma(gg\rightarrow t\bar{t})_{n=1}
&=& \int_{-1}^{1} \frac{ d\cos \theta }{32 \pi s} |\mathcal{M}(gg\rightarrow t\bar{t})|^2 \nonumber \\
&=& \frac{7}{240} \frac{g^4M_s^2}{32\pi} \left[ \frac{0.24}{(s-M_s^2)^2+(\Gamma_{g^*}^{J=2}M_s)^2} + \frac{0.76}{(s-M_s^2)^2+(\Gamma_{C^*}^{J=2}M_s)^2} \right]. ~~
\end{eqnarray}

\subsection{$n=2$ Resonances}

\subsubsection{Amplitudes}

The Veneziano amplitude near the second Regge resonance is approximated by expanding around $s=2M_s^2$ 
as 
\begin{eqnarray}
V(s,t,u)= \frac{\Gamma(1-\frac{u}{M_s^2})\Gamma(1-\frac{s}{M_s^2})}{\Gamma(1+\frac{t}{M_s^2})} \approx \frac{u(u+M_s^2) }{M_s^2(s-2M_s^2)} . 
\label{eq:V}
\end{eqnarray}
Therefore, 
we find 
 \begin{eqnarray}\label{equampl}
  |\mathcal{M}(gg\rightarrow t\bar{t})|^2  = \frac{3}{8}g^4M_s^4 \frac{(1-\cos \theta)^2 +(1+\cos \theta)^2 }{(s-2M_s^2)^2} \left(\frac{\sin \theta \cos \theta }{2}\right)^2 .
 \end{eqnarray}

The angular distribution can be expressed again in terms of the Wigner  $d$-functions as follows
\begin{eqnarray}
\frac{1-\cos(\theta)}{2}\frac{\sin(\theta)\cos(\theta)}{2} &=& \frac{1}{3}\sqrt{\frac{2}{5}}d^3_{2,1}(\theta)+\frac{1}{6} d^2_{2,1}(\theta), \nonumber \\
\frac{1+\cos(\theta)}{2}\frac{\sin(\theta)\cos(\theta)}{2} &=& \frac{1}{3}\sqrt{\frac{2}{5}}d^3_{2,-1}(\theta)-\frac{1}{6} d^2_{2,-1}(\theta).
\end{eqnarray}
Rewriting (\ref{equampl}) in Breit-Wigner form, 
\begin{eqnarray}
  &&|\mathcal{M}(gg\rightarrow t\bar{t})|^2  = g^4M_s^4[
                   \frac{1}{15} ~\frac{|d^3_{2,1}(\theta)|^2+|d^3_{2,-1}(\theta)|^2}{(s-2M_s^2)^2+2(\Gamma_{g^{**}}^{J=3}M_s)^2} \nonumber \\
                  && +\frac{1}{24} ~\frac{|d^2_{2,1}(\theta)|^2+|d^2_{2,-1}(\theta)|^2}{(s-2M_s^2)^2+2(\Gamma_{g^{**}}^{J=2}M_s)^2}
                   +\frac{1}{6}\sqrt{\frac{2}{5}} ~\frac{d^3_{2,1}(\theta)d^2_{2,1}(\theta)-d^3_{2,-1}(\theta)d^2_{2,-1}(\theta)}{(s-2M_s^2)^2+\Gamma_{g^{**}}^{J=3}\Gamma_{g^{**}}^{J=2}(2M_s^2)}] .
 \end{eqnarray}
There are two string resonances of spin $J=2,\ 3$ propagating in the $s$-channel. 
The third term on the right-hand side is an interference term between $J=2$ and $J=3$ resonances.
 In Fig.~\ref{figure4}(b) we plot the angular distribution. It is a superposition of P- and  D-waves. The interference
 vanishes upon integration and the cross section in the center of mass frame is 
\begin{eqnarray}
\nonumber 
\sigma(gg\rightarrow t\bar{t})_{n=2}
= \frac{g^4M_s^2}{32\pi} \left[ \frac{1}{60}\frac{1}{(s-2M_s^2)^2 +2(\Gamma_{g^{**}}^{J=2}M_s)^2} 
+ \frac{2}{105}\frac{1}{(s-2M_s^2)^2+2(\Gamma_{g^{**}}^{J=3}M_s)^2} \right] ,
\end{eqnarray}
where $\Gamma_{g^{**}}^{J=2}$ and $\Gamma_{g^{**}}^{J=3}$ are the decay widths of the $n=2$ string resonances with spin $J=2$ and $J=3$ . The decay rates to Standard Model particles are calculated in the following subsection, and can be found in Table \ref{decayratetable1}. The total decay rate should also include the decay channel to an $n=1$ string resonance plus Standard Model particles. Though we have not calculated this later decay rate in detail here, we argue below that it should be comparable to the decay rates to
a pair of Standard Model particles.

\subsubsection{The decay rate of an $n=2$ string resonance into Standard Model particles} \label{appendixdecayrate} 

We calculate some decay rates of an $n=2$ string resonance into Standard Model particles such as gluons and quarks, following the approach in  \cite{AGT3}. The Veneziano amplitude can be expanded around the second pole as 
in Eq.~(\ref{eq:V}).

We first look at the 4-gluon amplitudes. Using the approach in \cite{AGT3}, the decay width of $g^{**}$ resonance with spin $J$ and color index $a$ into two gluons with helicities and color indices $\lambda_1, \lambda_2$, $a_1,a_2$ can be written as 
\begin{eqnarray}
\Gamma^{aJ}_{\lambda_1 \lambda_2; a_1a_2} = \frac{1}{16(2J+1)\sqrt{2} \pi M_s} |F^{aJ}_{\lambda_1\lambda_2; a_1a_2}|^2 
\end{eqnarray}
Here $F^{aJ}_{\lambda_1\lambda_2; a_1a_2}$ is the matrix element for the decay of a resonance $g^{**}$ with $J_z=\lambda_1-\lambda_2$ into the two gluons, and can be extracted from the 4-gluon amplitude as 
\begin{eqnarray}
\langle 34;\theta |\mathcal{M} | 12; 0\rangle  &=& \sum_{a,J} \langle 34;\theta |\mathcal{M}^{aJ} | 12; 0\rangle \nonumber \\
&=& (s-2M_s^2)^{-1} F^{aJ}_{\lambda_3\lambda_4;a_3a_4}F^{aJ}_{\lambda_1\lambda_2;a_1a_2} d^{J}_{\lambda_1-\lambda_2; \lambda_3-\lambda_4}(\theta) 
\end{eqnarray}
where $d^J$ is the Wigner $d$-function as before.

We expand the 4-gluon amplitude around $s=2M_s^2$
\begin{eqnarray} 
 \mathcal{M}(g_1^{-}, g_2^{-}, g_3^{+}, g_4^{+})
&= &4g^2  \left[  \frac{s}{u} V_t ~ \Tr(T^{a_1}T^{a_2}T^{a_3}T^{a_4}+T^{a_2}T^{a_1}T^{a_4}T^{a_3})  
\right. \nonumber \\
 && \quad + \frac{s}{t} V_u ~\Tr( T^{a_2}T^{a_1}T^{a_3}T^{a_4}  
 + T^{a_1}T^{a_2}T^{a_4}T^{a_3})       \nonumber \\
 && \left. \quad +  \frac{s^2}{tu} V_s ~\Tr( T^{a_1}T^{a_3}T^{a_2}T^{a_4} + T^{a_3}T^{a_1}T^{a_4}T^{a_2}) \right] \nonumber \\
&=&  \frac{8g^2M_s^2 \cos(\theta)}{s-2M_s^2} \Tr( [T^{a_1},T^{a_2}] [T^{a_3},T^{a_4}])       \nonumber \\
&=& -\frac{8g^2 M_s^2}{s-2M_s^2} d^1_{0,0}(\theta) f^{a_1a_2a}f^{a_3a_4a}
\end{eqnarray}
Here $f^{a_1a_2a}$ is the antisymmetric $SU(N)$ structure constant. The color index $a=0$ corresponds to the $U(1)$ gauge boson $C$ in the $U(N)$, and since $f^{a_1a_20}=0$ we see that the $n=2$ resonance $C^{**}$ is not produced in gluon scattering and has no decay channel into two gluons. This feature of $n=2$ string resonances is different from that of the lowest $n=1$ resonances studied in \cite{AGT3}. 

From the above equation we see the string resonance has spin $J=1$ and up to a phase factor, the matrix element is 
\begin{eqnarray}
F^{a, J=1}_{++a_1a_2}=F^{a, J=1}_{-- a_1a_2}=2\sqrt{2} gM_s f^{a_1a_2a}
\end{eqnarray}
After taking into account a factor of $\frac{1}{2}$ from the double counting of identical particles, and also the fact that in this case these are two degenerate resonances with distinct  chiral properties, one of which decays into $(++)$ and the other one into $(--)$, we derive the decay width of the $J=1$ resonance into two gluons
\begin{eqnarray}
\Gamma^{J=1}_{g^{**}\rightarrow gg}= \frac{1}{2}\frac{1}{16\sqrt{2} (2J+1)\pi M_s} \sum_{a_1,a_2} |F^{a, J=1}_{++a_1a_2}|^2 
= \frac{g^2M_s} {12\sqrt{2} \pi} N
\end{eqnarray}
Here the color index $a\neq 0$, and we have used the $SU(N)$ Casimir invariant $C_2(N)=N$.

The other partial amplitudes can be obtained using cross symmetry. For example, the partial amplitude $\mathcal{M}(g_1^{-},g_2^+,g_3^{-},g_4^+)$ is obtained from the  $ \mathcal{M}(g_1^{-}, g_2^{-}, g_3^{+}, g_4^{+}) $ above by exchanging $s$ and $t$ variables, $a_2$ and $a_3$ indices.  We again expand around the second pole $s=2M_s^2$,
\begin{eqnarray}
\mathcal{M}(g_1^{-},g_2^+,g_3^{-},g_4^+)
&=& -\frac{8g^2}{s-2M_s^2} \left( \frac{1+\cos \theta }{2} \right)^2\cos \theta ~ f^{a_1a_2a}f^{a_3a_4a} \nonumber \\
&=&  -\frac{8g^2}{s-2M_s^2} \left( \frac{1}{3}d^3_{2,2}(\theta)+\frac{2}{3}d^2_{2,2}(\theta) \right)  f^{a_1a_2a}f^{a_3a_4a}
\end{eqnarray}
We see there are spin $J=2$ and $J=3$ resonances propagating in the s-channel. The matrix elements and decay widths are 
\begin{eqnarray}
&& F^{a,J=2}_{\pm\mp a_1a_2} = \frac{4}{\sqrt{3}} gM_s f^{a_1a_2a},
~~~~ F^{a,J=3}_{\pm\mp a_1a_2} = \frac{2\sqrt{2}}{\sqrt{3}} gM_s f^{a_1a_2a}
\\ && \Gamma^{J=2}_{g^{**}\rightarrow gg} = \frac{g^2 M_s}{15\sqrt{2}\pi} N, ~~~~ \Gamma^{J=3}_{g^**\rightarrow gg} = \frac{g^2 M_s}{42\sqrt{2} \pi } N
\end{eqnarray}

We then consider the decay channel to quarks 
\begin{eqnarray}
\mathcal{M} (q_1^{-}, \bar{q}_2^+,g_3^{-},g_4^+) 
= 4g^2M_s^2 \left( \frac{1}{3}\sqrt{\frac{2}{5}} d^3_{2,-1}(\theta) -\frac{1}{6} d^2_{2,-1}(\theta) \right) [T^{a_3},T^{a_4}]_{\alpha_1\alpha_2},
\end{eqnarray}
where $\alpha_1, \alpha_2$ are indices for quarks. We find the matrix elements and decay widths 
\begin{eqnarray}
&& F^{a,J=2}_{\pm\frac{1}{2}\mp\frac{1}{2}\alpha_1\alpha_2} = \frac{\sqrt{3}}{6} gM_s T^a_{\alpha_1\alpha_2},~~~~ 
F^{a,J=3}_{\pm\frac{1}{2}\mp\frac{1}{2}\alpha_1\alpha_2} = \frac{2}{\sqrt{15}} gM_s T^a_{\alpha_1\alpha_2} \\
&&  \Gamma^{J=2}_{g^{**}\rightarrow q\bar{q}} = \frac{g^2 M_s}{960 \sqrt{2}\pi} N_f, ~~~~ 
\Gamma^{J=3}_{g^{**}\rightarrow q\bar{q}} = \frac{g^2 M_s}{420 \sqrt{2}\pi} N_f
\end{eqnarray}

We list the decay widths of various channels in Table \ref{decayratetable1}. We see the decay width to quarks are much smaller than the decay width to gluons. In our analysis of the $t\bar{t}$ channel for the $n=2$ string resonances, we will encounter the spin $J=2$ and $J=3$ resonances, but not the $J=1$ resonance.   

\begin{table} 
\begin{center}
  \begin{tabular}{|c|c|c|c|}  
      \hline
    channel  & $\Gamma_{g^{**}}^{ J=1}$ & $\Gamma_{g^{**}}^{ J=2}$ &  $\Gamma_{g^{**}}^{ J=3}$  \\ \hline
    $gg$ & $ \frac{N} { 3 \sqrt{2} } $  & $\frac{4N}{15\sqrt{2}} $  & $ \frac{2N}{21\sqrt{2} } $  \\ \hline
     $q\bar{q}$ & 0 &  $\frac{N_f}{240 \sqrt{2}} $  &  $\frac{N_f}{105 \sqrt{2}} $  \\
    \hline
    \end{tabular}
    \end{center}
   \caption{The decay widths of $n=2$ string resonances. All quantities are to be multiplied by the factor $\frac{g^2}{4 \pi}M_s$.  For the Standard Model,  $N=3$, $N_f=6$. } \label{decayratetable1}
\end{table}

\subsubsection{Estimation of the total decay width of $n=2$ string resonance}

It is well known that the highest spin of string excitations
at level $n$  is $J=n+1$. Therefore, for an $n=2$ string resonance, there are three possible spins associating with it, listed in Table \ref{decayratetable1}, 
and the states of different $J$ should correspond to different particles, as in the case of the Standard Model.
Thus we could estimate the decay widths for the different $J$ states separately.

In section \ref{appendixdecayrate}, we derived the decay widths for the various channels of an $n=2$ string resonance decaying into two Standard Model particles (both
identified as $n=0$ ground states). Other than these channels, the $n=2$ string resonance ($\mathrm{SR_2}$) could also decay into an $n=1$ string resonance ($\mathrm{SR_1}$) and an $n=0$ Standard Model particle ($\mathrm{SM}$). The decay width of these later  processes is what we would like to estimate here. 
We will leave an explicit calculation of such decay width for future work, as it is sufficient for our purpose to estimate and compare it with the decay width of $\mathrm{SR_2} \to \mathrm{SM} + \mathrm{SM}$.
We will assume that the decay matrix elements 
for $\mathrm{SR_2} \to \mathrm{SR_1} + \mathrm{SM}$ to be comparable with $\mathrm{SR_2} \to \mathrm{SM} + \mathrm{SM}$.
We then count the multiplicity $N$ of the possible
 decay channels and multiply $N$ by the typical $\mathrm{SR_2} \to \mathrm{SM} + \mathrm{SM}$ widths  to get an estimate of these partial  widths (and thus eventually the total width).

Although there are many excited string states, most of them are not charged under the Standard Model and their presence does not concern us since they 
 will not be produced on resonance. 
Meanwhile, we can make a physically justified assumption that
the $n=0$ states do not contain non-Standard Model fields charged under the Standard Model gauge group\footnote{This means we assume that there are no chiral exotics charged under the Standard Model, and any vector-like states are made massive by whatever mechanism that stabilizes the moduli.}
(implicitly assumed in
 \cite{Cullen:2000ef,Cornet:2001gy,Friess:2002cc}) or else these new particles would have been observed in experiments.
 In principle, the $\mathrm{SR_1}$ states can decay to SMs with internal indices (e.g. a gluon in higher dimensions appear to be a scalar in 4-D) but such states are absent by assumption, so we only need to consider $\mathrm{SR_1}$ states with 4-D indices.

Take $n=2\ J=3$ string resonance ($\mathrm{SR_2^{J=3}}$) as an example. It could decay to either a pair of fermionic states or a pair of bosonic states. 
For the bosonic pair (i.e.~a 
 $\mathrm{SR_1}$ with integer spin ($J=1,2,3$) $ + $ a gluon), at $n=1$ level, we know there are 5 string resonances of gluons (see Appendix \ref{physdof}),  and each one corresponds to a particle, i.e. consisting of a spin 2 particle, two spin 1 particles and two spin 0 particles.  Since there are 5 $n=1$ particles, in the worst scenario, the multiplicity is 5. 
 However this is very unlikely, since the two $\mathrm{SR_1^{J=1}}$
 have different chiralities, so a  $\mathrm{SR_2}$
 usually decays to only one of them. The same argument applies to two $\mathrm{SR_1^{J=0}}$.
So the multiplicity is at most 3. Based on this argument and Table \ref{decayratetable1}, 
the decay width of  $\mathrm{SR_2^{J=3}}$ $\to$ integer spin $\mathrm{SR_1}$ $+$ gluon  should be roughly  $\mathrm{60\ GeV (M_S/1\ TeV)}$.

For the fermionic pair (i.e.~a $\mathrm{SR_1}$ with half-integer spin ($J=3/2, 1/2$) and a fermion), there are at most 4 $J=3/2$ resonances, and 2 $J=1/2$ resonances. So the number of decay channels is at most 6 times bigger than that of $\mathrm{SR_2}$ $\to q\bar{q}$. Then the decay width of these channels should be no more than 
$\mathrm{25\ GeV (M_S/1\ TeV)}$.

Collectively, the total decay width of $\mathrm{SR_2^{J=3}}$ is roughly 
$\mathrm{100\ GeV (M_S/1\ TeV)}$. 
On account of the same assumptions and method, the decay widths of $\mathrm{SR_2^{J=2}}$ and $\mathrm{SR_2^{J=1}}$ are roughly 
$\mathrm{250\ GeV (M_S/1\ TeV)}$ and $\mathrm{300\ GeV (M_S/1\ TeV)}$, respectively.
Since $J=3$ resonance has the narrowest width, it will be the dominant $\mathrm{SR_2}$ signal.

\section{$t\bar{t}$ Final State And String Resonances at the LHC}\label{Results}

\begin{figure}[tb]
  \begin{center}
    \begin{tabular}{cc}
      \resizebox{80mm}{!}{\includegraphics{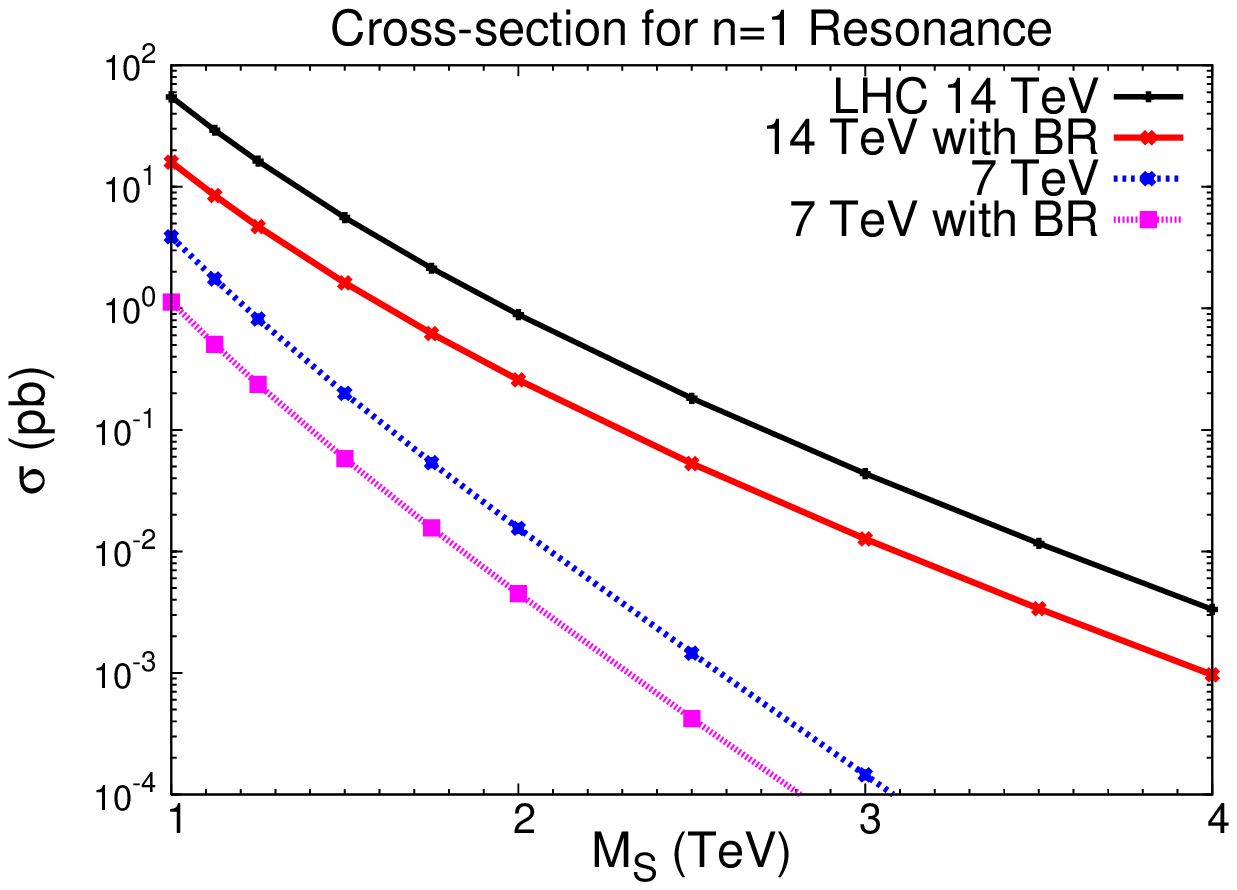} }
            \resizebox{80mm}{!}{\includegraphics{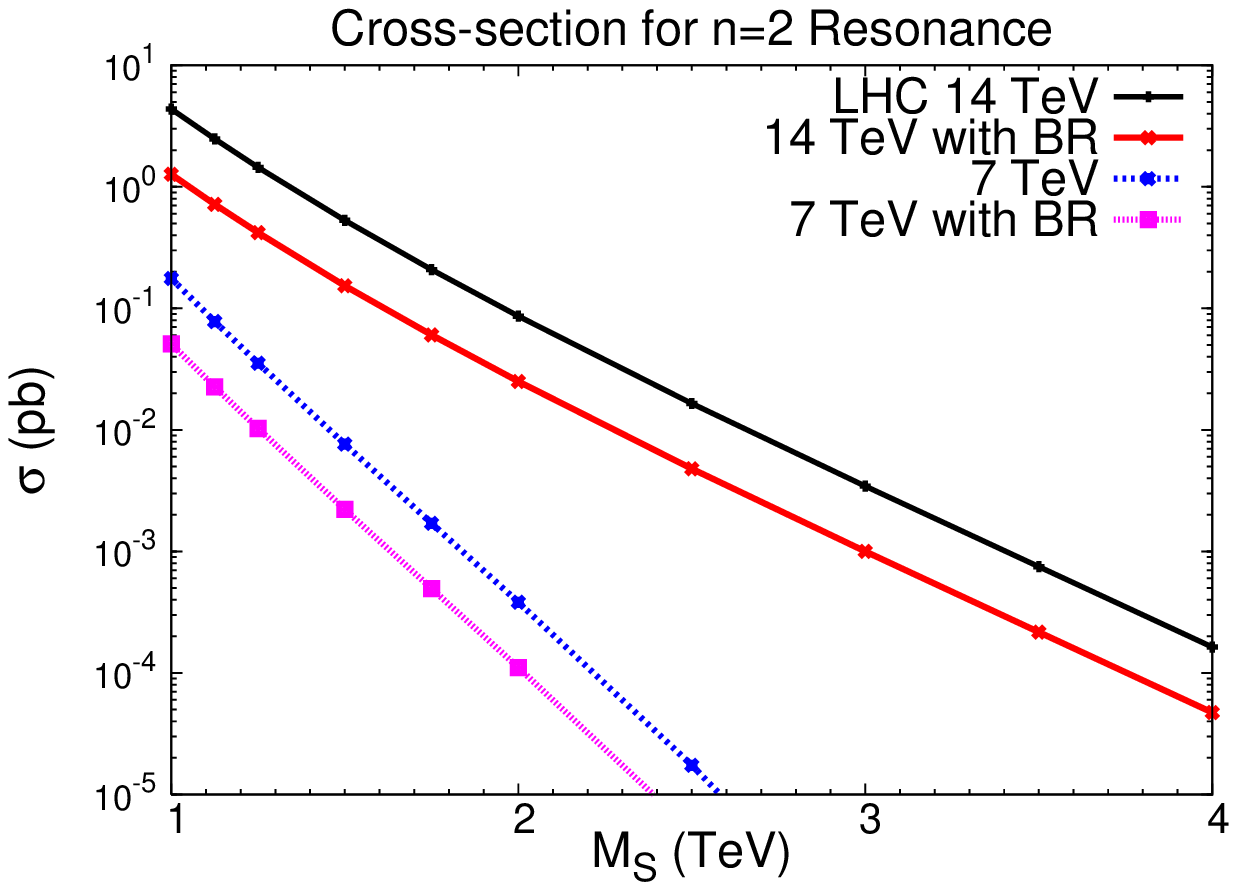} }
    \end{tabular}
    \caption{Total cross section for a string resonance production at the LHC 
versus it mass scale $M_{s}$ for (a) $n=1$ and (b) $n=2$ (mass 1.4$M_{s}$).
Solid curves represent $gg\to t\bar t$. Dashed curves include
the top quark decay branching fractions for the semi-leptonic mode $t \bar t \to b \ell \nu,\ b jj'$. }
    \label{fig:sigmatot}
  \end{center}
\end{figure}

As a top factory, the LHC will produce more than 80 million pairs of top quarks annually at the designed high luminosity  \cite{Han:2008xb}, largely due to the high gluon luminosity at the initial state.  It is therefore of great potential to observe string resonances in the $t\bar{t}$ channel if they couple to the gluons strongly. The Chan-Paton coeffient is unsuppressed (compared to say $ZZ$ production),  and
the Standard Model background is several orders of magnitude lower than that of the di-jet signal. 
Furthermore, the top quark is the only SM quark that the fundamental properties such as the spin and charge
may be carried through with the final state construction, and thus may provide additional information on the
intermediate resonances. The total production cross sections for a $t\bar t$ final state via a string resonance
are shown in Fig.~\ref{fig:sigmatot} by the solid curves at the LHC for 7 and 14 TeV, respectively. The dashed curves 
include the semi-leptonic branching fraction. The signal rates are calculated near the resonance peak
with $\pm 10\% M_{s}$. We see that the rate can be quite high. With 100 fb$^{-1}$ integrated luminosity
at 14 TeV, there will be about 100 $t \bar t \to b \ell \nu,\ b jj'$ events for an $n=1$ string resonance of 4 TeV
mass, and a handful events for $n=2$ of 5.7 TeV mass.

\subsection{Invariant Mass Distribution of $t \bar t$ Events}

With the semi-leptonic decay of $t\bar{t}$ one can effectively suppress the other SM backgrounds \cite{ATLASCMS}
Furthermore, the $t\bar t$ events may be fully reconstructable with the kinematical constraints \cite{Barger:2006hm}.
Thus, we consider here the semi-leptonic final state
\begin{equation}
t\bar{t} \to b\bar{\ell}\nu, \bar{b}jj ,
\end{equation}
with a combined branching fraction about $30\%$, including $\ell =e,\ \mu$. The total signal production rates including this semi-leptonic branching fraction are plotted in Fig.~\ref{fig:sigmatot} by the dashed curves. 

\begin{figure}[tb]
  \begin{center}
    \begin{tabular}{cc}
      \resizebox{80mm}{!}{\includegraphics{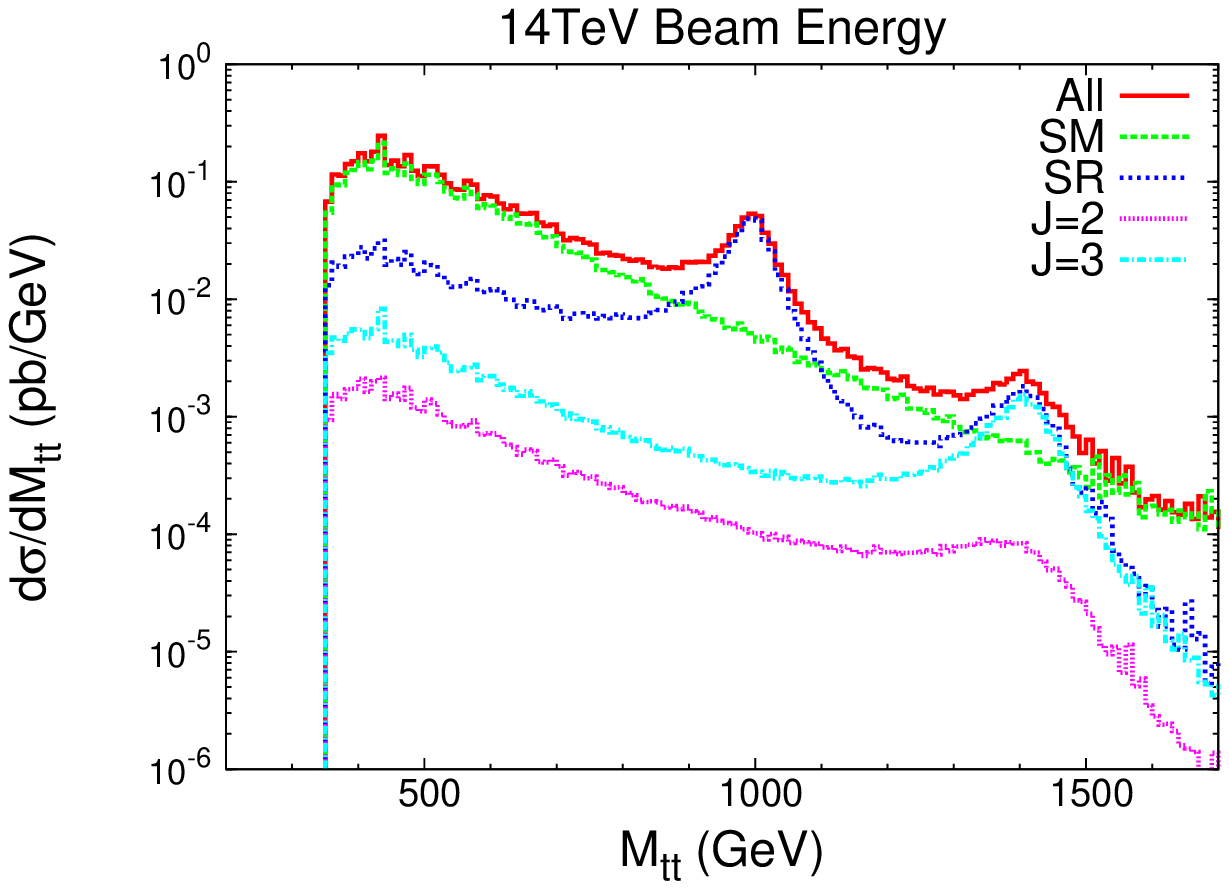}} &
      \resizebox{80mm}{!}{\includegraphics{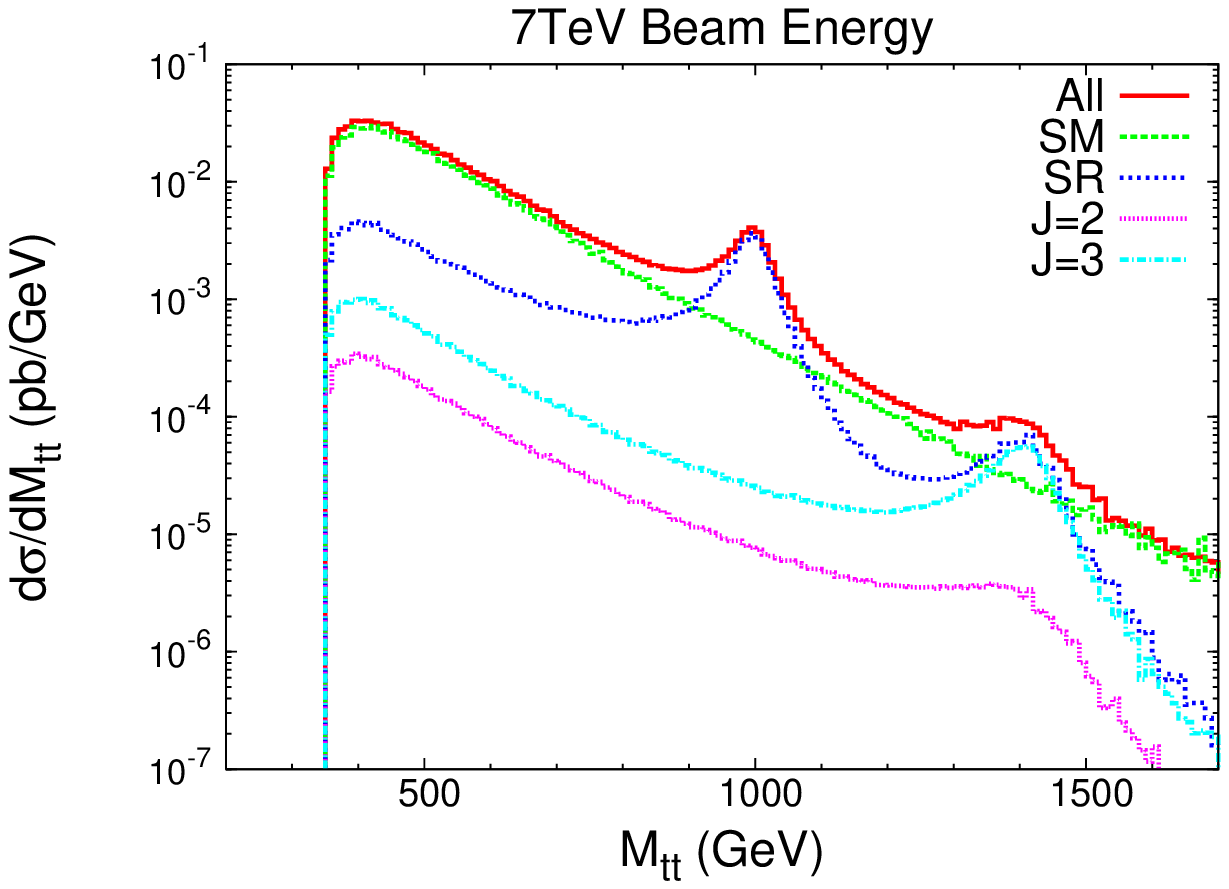}} \\
    \end{tabular}
    \caption{Invariant mass distribution of $t\bar{t}$ at the LHC 
    (a) with the c.m.~energy 14 TeV, 
    (b) with the c.m.~energy 7 TeV.
    Decay branching fractions with one hadronically 
    and the other leptonically have been included. 
    All channels (All), in the figure, include both string resonance 
    signal (SR) and Standard Model background (SM). 
    The distributions of $J=2$ and $J=3$ states are shown for  $n=2$ resonance, respectively.}
    \label{fig:mtt}
  \end{center}
\end{figure}

To simulate the detector effects, 
we impose some basic acceptance cuts on the momentum and rapidity of the final state leptons, 
jets and missing energy. We also demand the separation $\Delta R$ among the leptons and jets.
The cuts are summarized in Table \ref{tab:cut}.
Figure \ref{fig:mtt} shows that both $n=1$ (peak position at $M_{S}$=1 TeV)
and $n=2$ (peak position at $M_{S}$=1.41 TeV) string resonances are evident at LHC with the c.m.~energy at 14 TeV and 7 TeV.  

\begin{table}[tb]
  \caption{Basic acceptance cuts for $t\bar{t}$ events at the LHC.}
    \centerline{\begin{tabular}{@{}c|c|c@{}}
      \hline
      \hline
      &$p_{T}$ (GeV)& $\eta$ rapidity \\
      \hline
      ~~$\ell$~ & 20 & 3  \\
      \hline
      ~~$j$~ & 30 & 3  \\
      \hline
      ~~$\etmiss$~&30& N/A\\
      \hline
      \hline
      ~~$\Delta R^{cut}$~&0.4&0.4\\
      \hline 
    \end{tabular}
}
  \label{tab:cut}
\end{table}

Though the Standard Model background is large as seen in the figure for the continuum spectrum, 
we still get abundant resonant events in the peak region.
Assuming the annual luminosity at the LHC to be $\mathrm{10^{34}~cm^{-2} s^{-1}\sim 100~fb^{-1}}$ per year 
with the c.m.~energy 14 TeV, one would expect  to have about a million $n=1$ 
string resonance events in the peak region 900-1100 TeV, for ${M_{s}}$=1 TeV
(with decay width calculated by \cite{AGT3}). 
Similarly, one may expect about a thousand events around ${m_{tt} \approx 1.4}$ TeV. 
At the lower energy of 7 TeV with an integrated luminosity 1 fb$^{-1}$ as the current planing for the initial 
LHC running, we would expect to have about 500 events near $m_{tt} =1$ TeV, 
and about one event at 1.4 TeV. 

A distinctive feature of string resonances is their mass ratio. Suppose the masses of the first and the second excited string states are $M_{t\bar{t}1}$, and $M_{t\bar{t}2}$ respectively,  then
$M_{t\bar{t}2}/M_{t\bar{t}1} = \sqrt{2} \approx 1.4$ (as in Figure \ref{fig:mtt}), which can potentially distinguish them
from other kinds of resonances.  
It is important to note that, at leading order of the gauge coupling,
 this mass ratio  is {\it model independent}
 and does not depend on the geometry of compactification, the configuration of branes, and the number of supersymmetries.
 Higher order corrections to this ratio depend on  model specific details, and if measurable,
  can serve as a discriminator of different string theory compactifications.

\subsection{Angular Distribution of $t\bar{t}$ Events}

  \begin{figure}
    \begin{center}
      \resizebox{80mm}{!}{\includegraphics{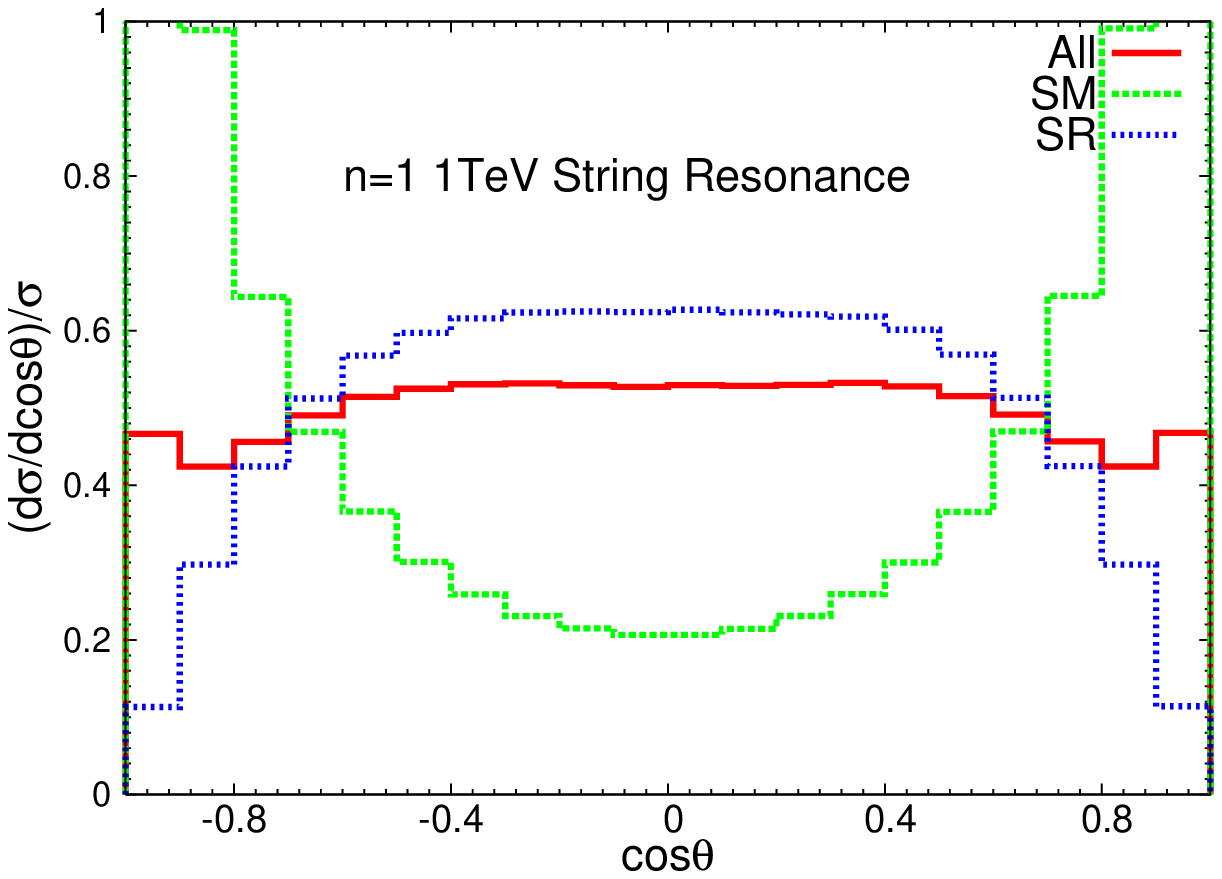}}
            \resizebox{80mm}{!}{\includegraphics{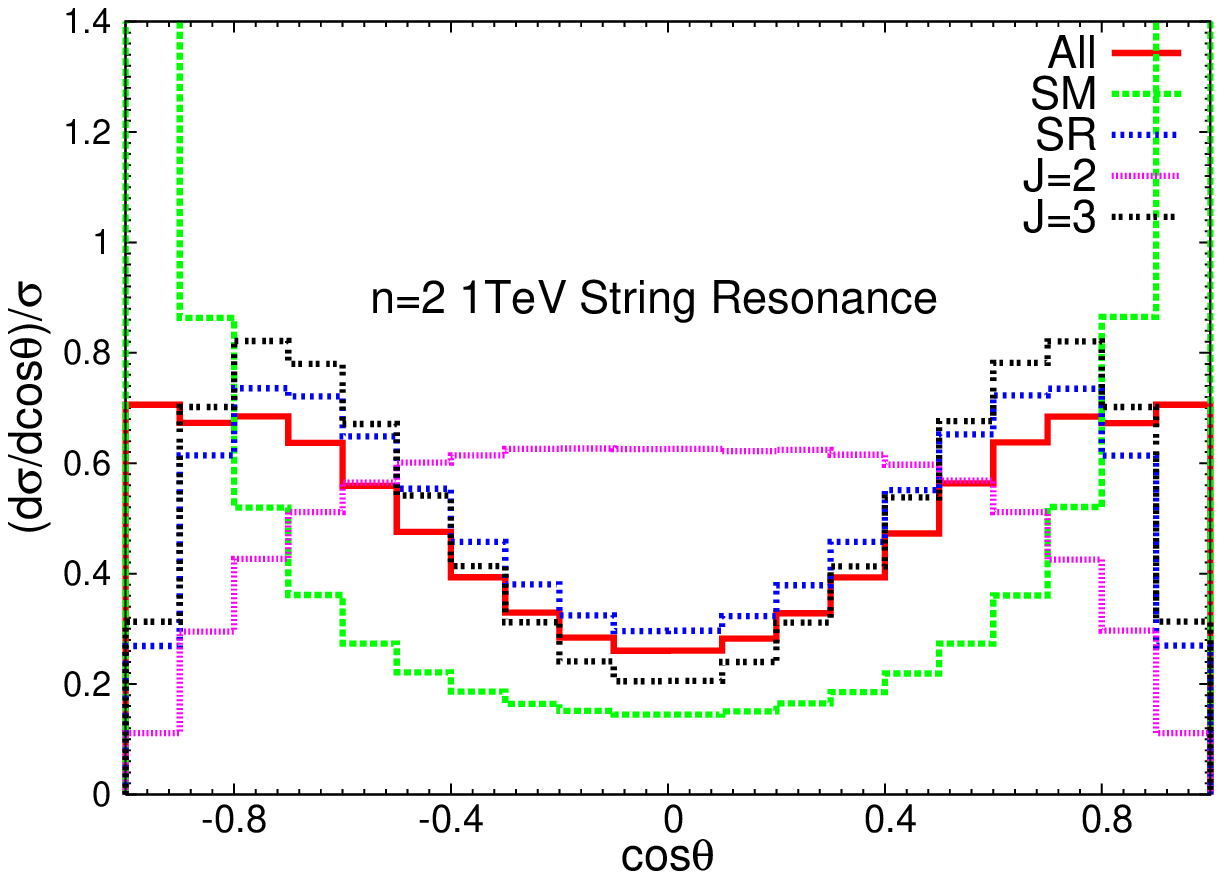}}
      \caption{Normalized angular distribution of $t\bar{t}$ at the LHC 
        via the string resonance exchange 
         in the peak region $\mathrm{900-1100~GeV}$ for $n=1$,
         and in the region $\mathrm{1350-1450~GeV}$ for $n=2$.}
      \label{fig:cos1}
    \end{center}
  \end{figure}

Other than the distinctive invariant mass distribution,
the normalized angular distribution
can 
provide a great discriminator between
 string resonances and
the Standard Model background.
The shape of the angular distribution for a string resonance is mainly a result of 
the Regge behavior of the string amplitudes. 
Figure \ref{fig:cos1} shows the normalized angular distributions 
according to the cross section of all channels 
(including string resonance and Standard Model $t\bar{t}$ background)
in the  $900-1100$ GeV peak region for $n=1$,
and in the $1350-1450$ GeV region for $n=2$.
The angle is defined for the outgoing top quark with respect to the incoming beam direction
in the $t\bar t$ c.m.~frame, which can be reconstructed for the semi-leptonic channel
on an event-by-event basis. 
We observe the qualitative difference between the string resonance signal and the SM background:
The signal is protrudent  at the large scattering angular region $\mathrm{\cos\theta \to 0}$, 
where the Standard Model background
is collimated with the beam direction, due to the dominant behavior
of the $t$- or $u$-channel, scaling as $(1\pm \cos\theta)^{-1}$. 
As shown earlier, there is even a clear difference in shape between the $n=1$ and $n=2$ states, as indicated by
the solid curves (All). 
It is interesting to note that for an $n=2$ resonance, the shape difference between $J=2$ and $J=3$ states 
is also evident. However, 
since they are all degenerate in mass, one would have to use more sophisticated fits to their angular distributions to disentangle them.

The major advantages for considering the semi-leptonic decay of the $t\bar t$ are that (1) we will be able to tag the
top versus anti-top, and (2) the angular distribution
of the charged lepton in the reconstructed c.m.~frame can carry  information about the top-quark spin
correlation \cite{Spin}, and thus provides an effective test for the nature of the top-quark coupling. 
Since the resonance couplings under our consideration are all vector-like, we will not pursue further detail studies of these variables. 

\section{$t\bar{t}g$ Final State And String Resonances at the LHC}

\begin{figure}
  \begin{center}
    \begin{tabular}{cc}
      \resizebox{80mm}{!}{\includegraphics{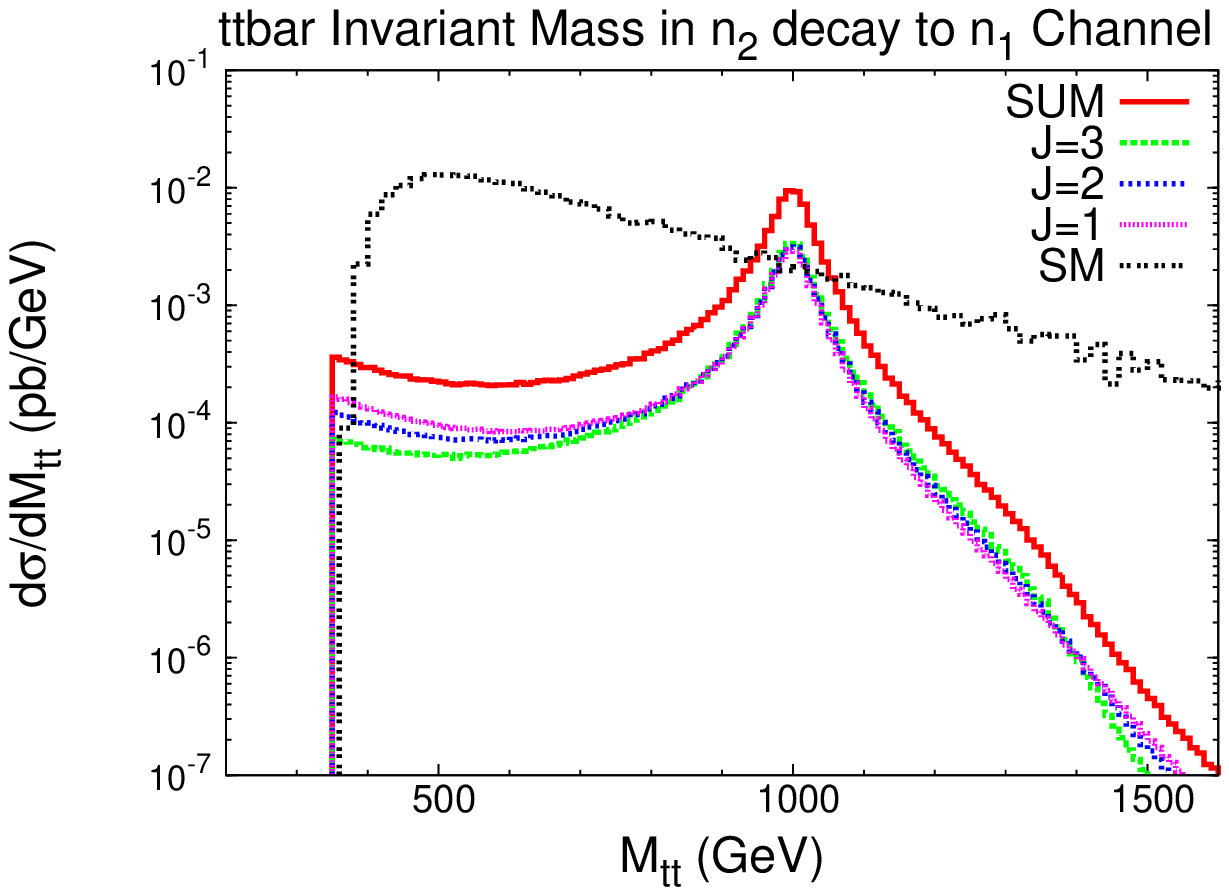}}&
      \resizebox{80mm}{!}{\includegraphics{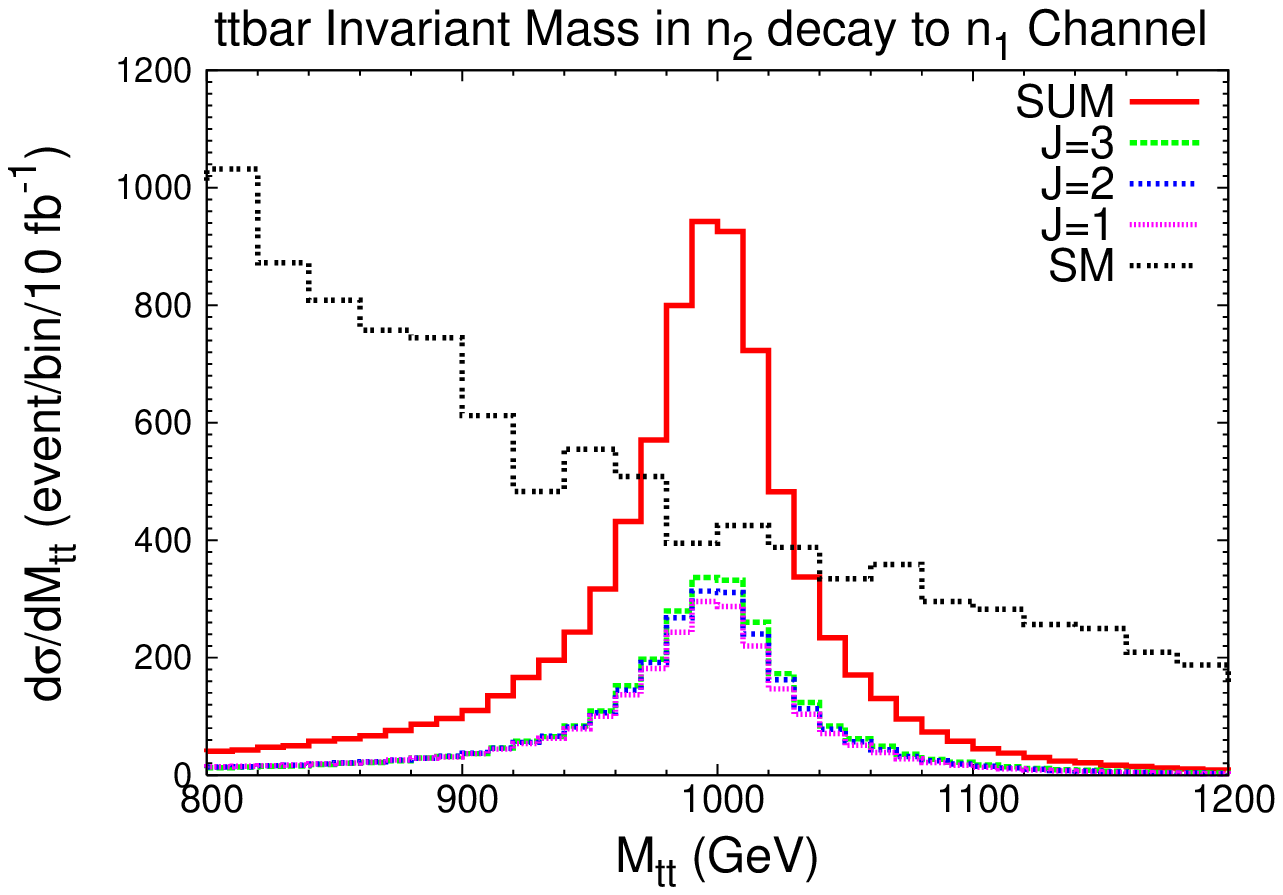}}\\
      \resizebox{80mm}{!}{\includegraphics{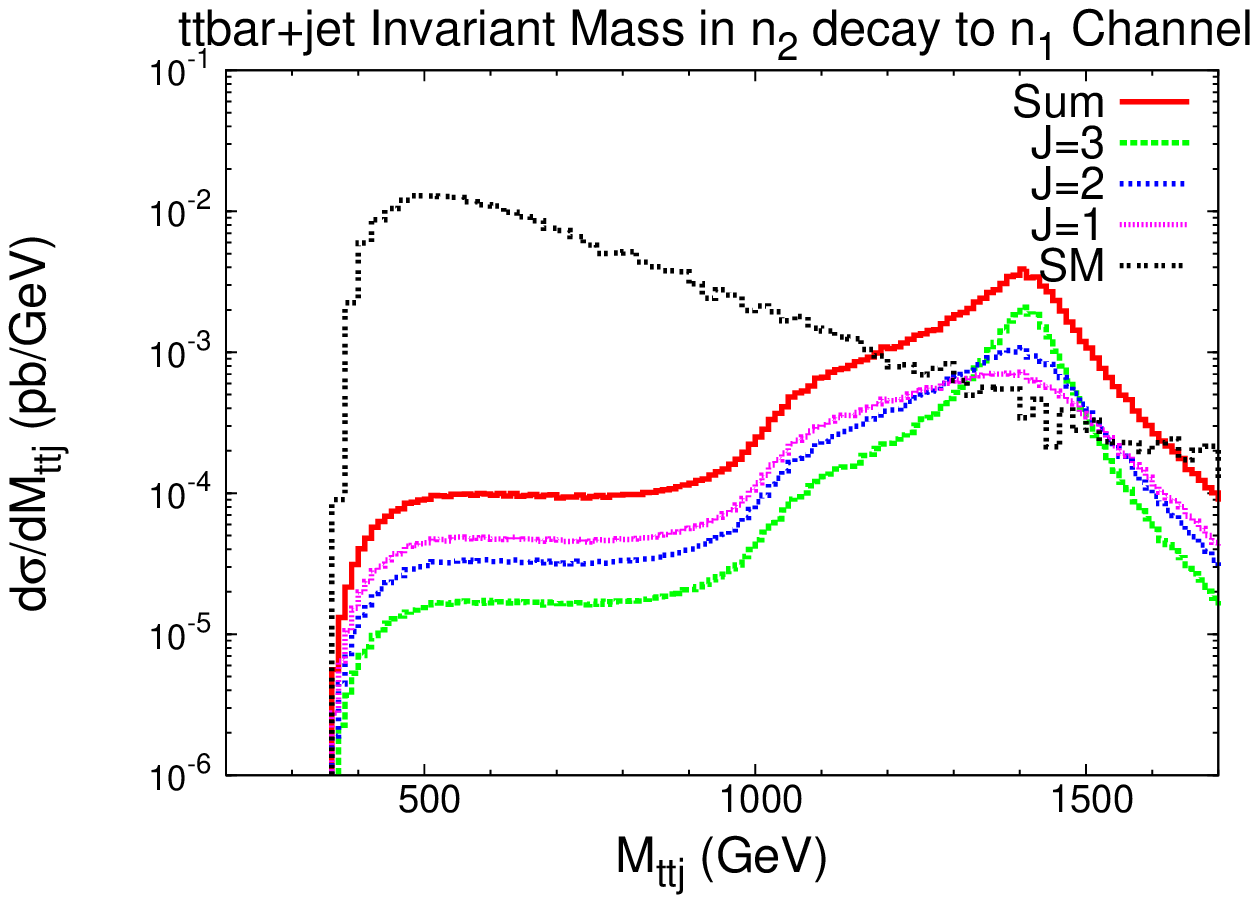}}&
      \resizebox{80mm}{!}{\includegraphics{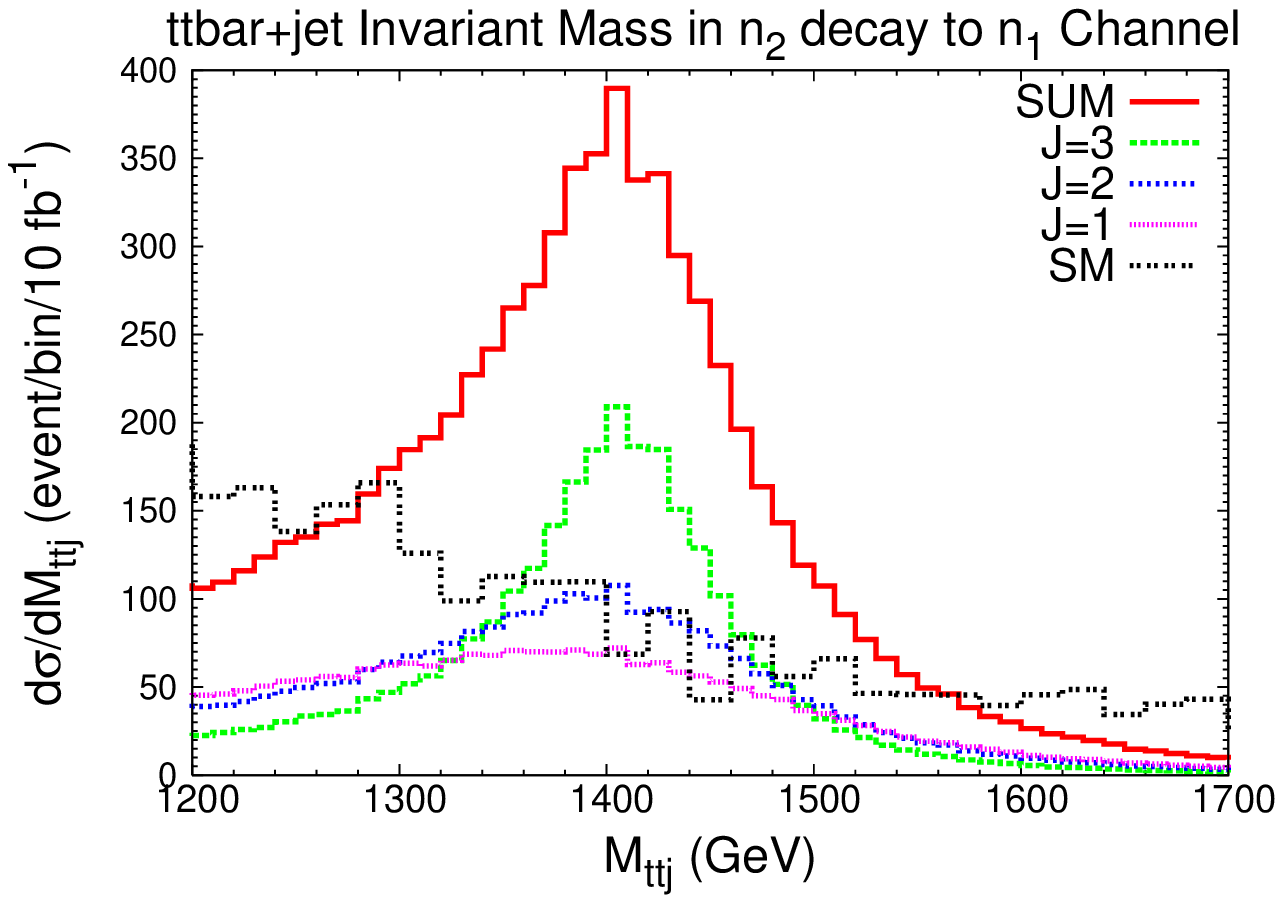}}\\      
    \end{tabular}
    \caption{
      Invariant mass distributions in the $g+g \to n=2$ resonance $\to n=1$ resonance + $g \to t\bar{t}+g$ channel.
      Upper two graphs: $t\bar{t}$ invariant mass signal and the Standard Model background. 
      Lower two graphs: $t\bar{t}+g$ invariant mass signal and the Standard Model background.
      The summed signal, in the figure, includes the contributions from both $J=2$ and $J=3$ resonances.} 
    \label{fig:mttj}
  \end{center}
\end{figure}

For an $n=2$ string resonance, it could decay not only directly to two $n=0$ Standard Model particles as discussed above, but also to an $n=1$ string resonance and an $n=0$ Standard Model particle. Thus, there is an additional decay chain 
\begin{equation}
(n=2) \to (n=1) + g \to t\bar{t} +g ,
\end{equation}
following an $n=2$ resonance production. 
The kinematics of this channel  may lead to further distinctive signatures. First of all, the additional gluonic jet
is highly energetic, with an energy of the order $M_{s}/\sqrt{2}$, 
quite distinctive from QCD jets. Secondly, there are two resonances with $n=1$
and $2$ respectively, both appearing in the same event sample.  One would thus expect to establish a convincing signal observation
with the mass peaks at $M_{tt}\sim M_{s}$ and $M_{ttj} \sim 1.4M_{s}$. 
Figure \ref{fig:mttj} shows the invariant mass distributions for $t\bar{t}g$ production from an $M_{s}=1$ TeV string resonance and the Standard Model background. The upper panels present the $M_{tt}$ distribution where an $n=1$ resonance peak is apparent. The lower panels are for  the $M_{ttj}$ distribution which shows the expected $n=2$ resonance peak. 
The two linear plots on the right column are the blow-up view near the resonances in  units of number of events
per bin (10 GeV) with an integrated luminosity of 10 fb$^{-1}$. The different angular momentum states $J=1,2,3$
are also separately plotted. Since they are all degenerate in mass, one would have to use more sophisticated fits to their angular distributions to disentangle them, as discussed in the previous sections. 

\section{Higher String Scale signal}\label{HighScale}

\begin{figure}
  \begin{center}
    \begin{tabular}{cc}
      \resizebox{80mm}{!}{\includegraphics{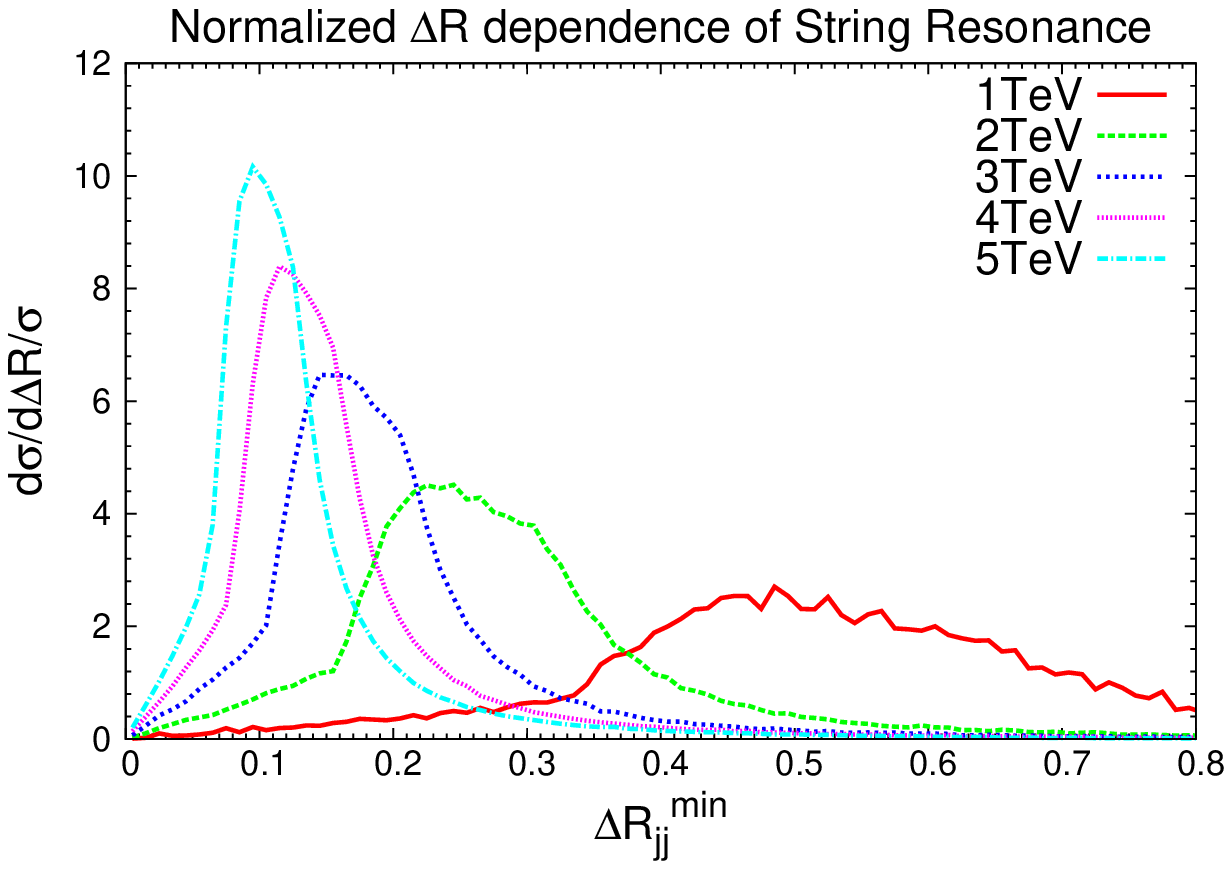}} &
      \resizebox{80mm}{!}{\includegraphics{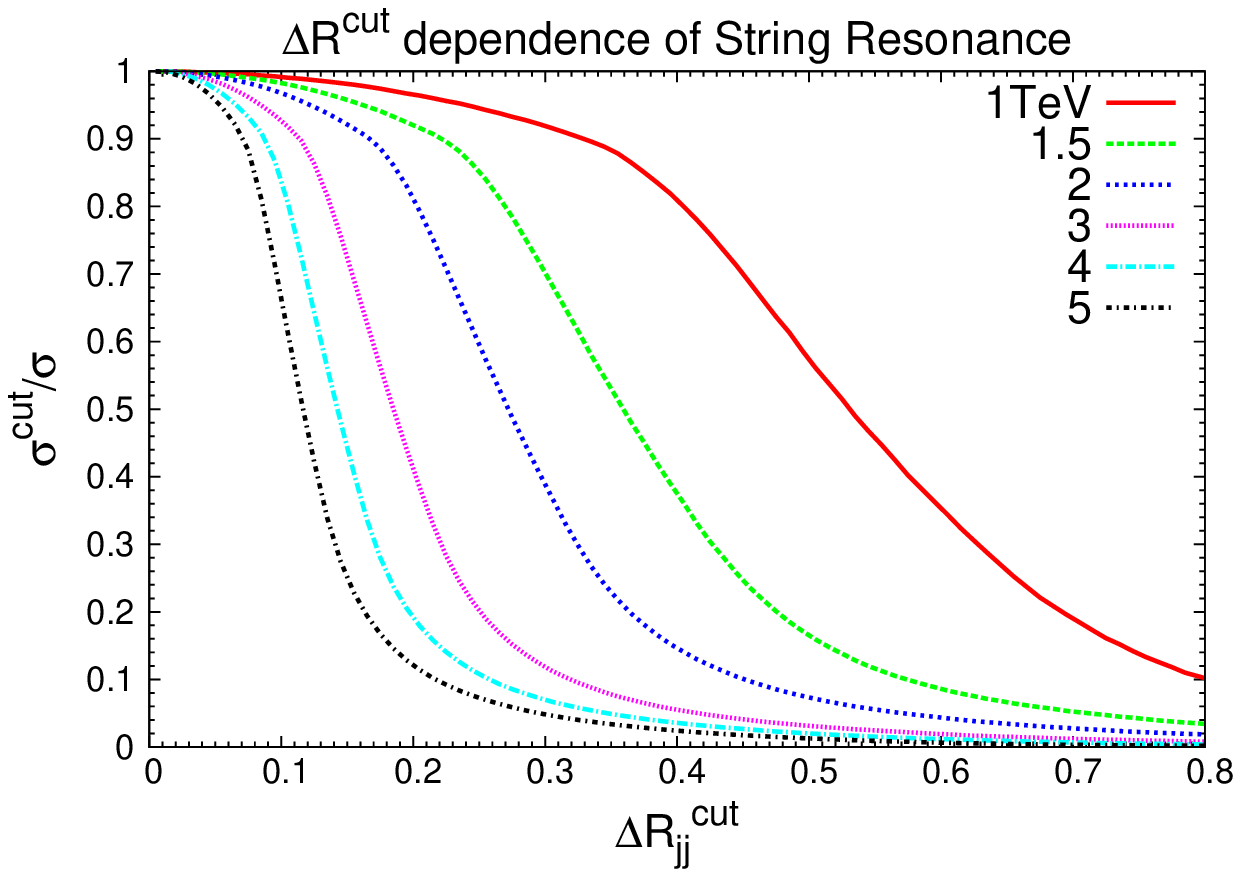}} \\
    \end{tabular}
    \caption{
    (a) The distribution of the minimum separation between any two jets  from $t(\bar{t})$ decay $\mathrm{\Delta R_{jj}}$.
    (b) $\mathrm{\Delta R_{jj}^{cut}}$ Efficiency: Peak region cross section percentage versus $\mathrm{\Delta R_{jj}^{cut}}$ of 
    $t\bar t$ semi-leptonic decay at the LHC. The energy labels in the legends stand for the resonance masses.}
    \label{fig:dRjj}
  \end{center}
\end{figure}

As seen from Fig.~\ref{fig:sigmatot}, a string resonance of mass about $4$ TeV  for both $n=1$ and 2
may be copiously produced at the LHC with the designed luminosity of 100 fb$^{-1}.$
However, as the mass of the string resonance increases, the decay products become more and more collimated,
and the fast moving top quark is a ``top jet''. 
Figure \ref{fig:dRjj} reveals the dependance of the cross section 
on the minimum separation between any two jets from top decay
in the peak region ($\pm$ 100 GeV around the peak).
Based on the figures, one notices that, for an  $1-1.5$ TeV string resonance,
$\mathrm{\Delta R^{cut}} = 0.3$ is efficient in defining an isolated lepton or jet.
For a higher string mass, the top quarks are highly boosted and too collimated to be identified as multiple jets. 
One may use various methods \cite{Kaplan:2008tt,Thaler:2008w} to identify the highly boosted top produced 
by string resonances as a single ``fat top jet''. Non-isolated muons from the top decay may still help
in  top-quark identification. 

Furthermore, the granularity of the hadronic calorimeter is roughly $\Delta \eta \times \Delta \phi \sim 0.1 \times 0.1$\cite{ATLASCMS}.
To identify individual decay products, 
one needs the separation $\mathrm\Delta R$ between any two decay products to be at least $0.1-0.2$.
From Figure \ref{fig:dRjj}, we see that we could
identify the boosted tops if the string scale is below $4$ TeV. 
For a string scale above $4$ TeV, 
we could not tell the difference between top quarks and other QCD jets from
light quarks or gluons. 
Then, our background will be from QCD dijet, and there would be no advantage in considering $t\bar{t}$ 
final state. The situation is similar to the study in \cite{AGLNST} for light quark final states. 

\section{Discussions and Conclusion}\label{Conclusion}

In this paper, we have studied the discovery potential of string resonances at the LHC via the $t\bar t$ final state.
In a large class of string models where the Standard Model fields are localized on the worldvolume of D-branes, the string theory amplitudes of 
certain processes at hadron colliders are {\it universal} to leading order in the gauge couplings.
This universality makes it possible to compute these {\it genuine} string effects which are independent of the geometry of the extra dimensions, the configuration of branes, and whether 
supersymmetry is broken or not.
Among the various processes, we found the production of
$t \overline{t}$ pairs at the LHC to be advantageous 
to uncover the properties of excited string states which appear as resonances in these amplitudes.
The top quark events are distinctive in event construction and have less severe QCD backgrounds than other light  jet signals from string resonance decays. 
The swift decay of top quarks via weak interaction before hadronization sets in makes it possible to 
reconstruct the full decay kinematics and to take advantage of its spin information.
We investigated the invariant mass distributions and the angular distributions of 
$t \overline{t}$ production which may signify the exchange of  string resonances.

Our detailed phenomenological studies suggest that string resonances can be observed in the
$t\bar{t}$ channel for a string scale up to 4 TeV by analyzing both the invariant mass distributions and the angular distributions.
For a string scale less than $1.5$ TeV, we proposed to use semi-leptonic decay to reconstruct the $t\bar{t}$ c.m.~frame and obtain the angular distributions to disentangle  string resonances with different angular momenta. 
For a string scale between $1.5$ TeV and $4$ TeV, we need to identify the boosted tops as a single top jet and directly observe the angular distribution. 
If the string scale is higher than $4$ TeV, we cannot observe the substructure of quark jets and so
 the signal will be submerged by the QCD dijet background.
The potential benefits for using the $t\bar t$ final state in the semi-leptonic mode 
is that one would be able to tag $t$ from $\bar t$, so that 
the coupling properties of the string resonance, such as parity, CP, and chirality, could be studied via the angular
distributions of the top, especially the distributions of the charged lepton with the help of the spin-correlation. 

The current work focuses on string
amplitudes which give the leading order (in string coupling) 
contribution to $q \overline{q}$ production at the LHC.
These amplitudes are model-independent as they do not involve exchange of 
KK modes. As a result, such amplitudes do not
distinguish between different quarks (their flavors or chiralities) even though they may have different 
wavefunctions (classical profiles) in the internal space.
This is however not the case for the subleading contributions to $q \overline{q}$ 
production 
as one finds KK modes propagating as intermediate states in these processes (e.g. in 4 fermion amplitudes).
The overlap of the KK mode wavefunctions with that of the quarks
determines the coupling strength of these processes. For example, in warped extra dimensional scenarios, the KK modes are localized in the highly warped (or the so called infrared) region so are the heavy quarks.
In these scenarios, the production of $t \overline{t}$ is dominant among such 4-fermion amplitudes.
Given that phenomenological constraints imply that string states are generically at most a factor of a few heavier than the lightest KK modes \cite{Reece:2010xj,Hassanain:2009at,Perelstein:2009qi} in Randall-Sundrum like scenarios, it is worthwhile to explore warped extra dimensional models in the context of string theory\footnote{Even within an effective field theory context, the kinds of warped geometries arising in string theory
have motivated new
model building possibilities, see, e.g., \cite{Shiu:2007tn,McGuirk:2007er}.}.
For example, the open string wavefunctions obtained in \cite{Marchesano:2008rg} and the effective action describing closed string fluctuations (massless and KK modes) in warped compactifcations 
\cite{Shiu:2008ry,Douglas:2009zn}\footnote{For some earlier work discussing issues in the effective theory of warped string compactifications, see e.g., \cite{DeWolfe:2002nn,Giddings:2005ff,Burgess:2006mn}.}
may be useful in determining the aforementioned amplitudes.
Furthermore, quarks with different chiralities can attribute differently to these amplitudes, as they have a different origin in D-brane constructions. The left-handed quarks are doublets under the weak interaction and so 
the corresponding open strings end on the weak $SU(2)$ branes
whereas those associated with the right-handed quarks do not.
Thus, e.g., in the Drell-Yan process $q\bar{q}\rightarrow \ell\bar{\ell}$, certain parts of the amplitudes are only attributed to the left-handed quarks \cite{LST}. 
Similar features are expected for processes involving quark final states
such as $q \overline{q} \rightarrow t \overline{t}$.
It would be interesting to study whether measurements of these chiral couplings can be used to distinguish between different string theory models.

\vspace{0.2in} {\leftline {\bf Acknowledgments:}}

\bigskip

We thank Diego Chialva, Michael Ramsey-Musolf for discussions and comments.
This work was supported in part by a DOE grant DE-FG-02-95ER40896.
GS  was also supported in part by NSF CAREER Award No.~PHY-0348093, a Cottrell Scholar Award from Research
Corporation, a Vilas Associate Award from the University of Wisconsin, and a John Simon Guggenheim Memorial Foundation Fellowship. GS would also like to acknowledge the hospitality of the Hong Kong Institute for Advanced Study durng part of this work.

\bigskip \bigskip

\appendix 

\centerline{\Large \bf Appendices}
\section{Physical Degrees of Freedom Counting for $n=1$ Resonance} \label{physdof}

Here we focus on gluons and their excited string modes. In D-brane model building, the gluons are realized as open strings attached to a stack of 3 D-branes, forming adjoint representation of a $U(3)$ gauge group. The gluons are represented by a vertex operator:
\begin{eqnarray}
T^a e_\mu\psi^\mu_{-\frac{1}{2}}|0;k\rangle
\end{eqnarray}
where $T^a$ is the Chan-Paton matrix,  $e_\mu$ is the polarization vector of the gluon, $\psi_{-\frac{1}{2}}$ is a world-sheet fermion creation operator, and $|0;k\rangle $ is the open string vacuum state in the NS sector.  Here we consider string states in 4 dimension, so the index $\mu$ goes over $0,1,2,3$ and the momentum $k$ is a 4-dimensional momentum. We will use the $(-,+,+,+)$ signature. The gluons are massless, and are the lowest string states because the NS vacuum is projected out by the GSO projection. The physical state conditions constrain the momentum $k^2=0$ so this is a massless vector particle. The polarization vector also satisfies the physical state condition $e\cdot k=0$ with the equivalence condition $e\cong e+k$.  

We can write the vertex operator of gluons using the state-operator correspondence. For open strings, we replace the bosonic and fermionic creation operators with world-sheet bosons and fermions as  follows:
\begin{eqnarray} \label{stateoperator}
\alpha_{-m}^\mu &\rightarrow & i (\frac{1}{2\alpha^\prime})^{\frac{1}{2}}\frac{1}{(m-1)!}\partial ^m X, \nonumber \\
\psi_{-r}^\mu &\rightarrow & \frac{1}{(r-\frac{1}{2})!} \partial ^{r-\frac{1}{2}}\psi ^\mu
\end{eqnarray}
The vertex operators for gluons in the $-1$ and $0$ pictures are the following:
\begin{eqnarray}
\mathcal{O}(z)_{-1} &=& T^a e_{\mu}e^{-\phi}\psi^\mu(z) e^{ik\cdot X(z)}, \nonumber \\
\mathcal{O}(z)_0 &=& T^a e_\mu [i\partial X^\mu + (2\alpha^\prime) (k\cdot \psi) \psi^\mu] e^{ik\cdot X(z)}
\end{eqnarray}

Now we consider the next level of string states. Because of the GSO projection, the next level number is $\frac{3}{2}$. They will be referred to as the first excited string modes (or $n=1$ string modes), and we refer to the gluons as $n=0$ string modes. Using the bosonic and fermionic creation operators, a general $n=1$ open string state  can be written as 
\begin{eqnarray} \label{stringstate}
|\chi\rangle = [(\xi_1)_\mu \psi_{-\frac{3}{2}}^\mu+(\xi_2)_{\mu\nu}\psi_{-\frac{1}{2}}^\mu\alpha_{-1}^\nu+(\xi_3)_{\mu\nu\rho}\psi_{-\frac{1}{2}}^{\mu}\psi_{-\frac{1}{2}}^{\nu}\psi_{-\frac{1}{2}}^{\rho}] |0;k\rangle
\end{eqnarray}
where $(\xi_3)_{\mu\nu\rho}$ is antisymmetric since the fermionic operators $\psi_{-\frac{1}{2}}$ anti-commute.  

We will use the OCQ (old covariant quantization) method for quantizing  string theory, as we only need to deal with the matter sector in this formulation. The physical state conditions for $|\chi\rangle$ are as follows: 
\begin{eqnarray} \label{physicalstatecondition}
(L_0-\frac{1}{2})|\chi\rangle=0, ~~ L_1|\chi\rangle =0, ~~ G_{\frac{1}{2}}|\chi\rangle =0, ~~ G_{\frac{3}{2}}|\chi\rangle =0 
\end{eqnarray}
Here the $L_m$ and $G_r$ are superconformal Virasoro generators for the matter sector on the world-sheet, and the constant $-\frac{1}{2}$ in the first equation takes into account the contributions from the ghost sector in the NS vacuum. The superconformal Virasoro generators are: 
\begin{eqnarray}
L_m &=& \frac{1}{2}\sum_{n\in \mathcal{Z}}: \alpha^\mu_{m-n}\alpha_{\mu n}: 
+\frac{1}{4}\sum_{r\in \mathcal{Z}+\frac{1}{2}}(2r-m):\psi^\mu_{m-r} \psi_{\mu r}: + a\delta_{m0}, \nonumber \\
G_r &=& \sum_{n\in \mathcal{Z}}\alpha^\mu_n(\psi_\mu)_{r-n} 
\end{eqnarray}
Here the constant $a=0$ in the NS sector, and the zero mode bosonic generator is related to the momentum $\alpha_0^\mu= (2\alpha^\prime)^{\frac{1}{2}}k^\mu$ for the open string sector we consider.  The $::$ denotes normal ordering of creation operators with negative indices and annihilation operators with positive indices. 

The zero mode of the Virasoro generator is $L_0=\alpha^\prime k^2+N$ where $N=\frac{3}{2}$ is the level number for the string state $|\chi\rangle$ in (\ref{stringstate}), so the the first physical state condition in (\ref{physicalstatecondition}) gives $k^2=-\frac{1}{\alpha^\prime}$. The mass of the $n=1$ string mode $|\chi\rangle$ is 
\begin{eqnarray}
 m=
 \frac{1}{\sqrt{\alpha'}} \equiv M_s
 \end{eqnarray} 
 
 Using the commutation relation of the bosonic and fermionic operators $\{\psi^\mu_r,  \psi^\nu_s\}=\eta^{\mu\nu}\delta_{r,-s}$, $[\alpha^\mu_m,\alpha^\nu_n]=m\eta^{\mu\nu}\delta_{m,-n}$, we find that the rest of the physical state conditions in (\ref{physicalstatecondition}) are given by:
 \begin{eqnarray} \label{constrain}
 	(2\alpha^\prime)^{\frac{1}{2}}(\xi_1)\cdot k+(\xi_2)_{\mu\nu} \eta^{\mu\nu} &=& 0 \nonumber \\
 	(\xi_1)_\nu + (2\alpha^\prime) ^{\frac{1}{2}}k^\mu(\xi_2)_{\mu\nu} &=& 0 \nonumber \\
 	 (\xi_2)_{\mu\nu}-(\xi_2)_{\nu\mu}+6(2\alpha^\prime)^{\frac{1}{2}}k^\rho(\xi_3)_{\mu\nu\rho} &=& 0 
\end{eqnarray}

We also need to consider possible ``null states''  which are states that can be written as $L_n|\phi\rangle$ or $G_r|\phi\rangle$ with $n,r>0$ for any state $|\phi\rangle$. It turns out that at this level $\frac{3}{2}$, all physical states satisfying the constrains (\ref{constrain}) are not null. An intuitive understanding is that because the $n=1$ states are massive,  all physical polarizations are non-trivial, as opposed to the case of massless vector particle where one does not have longitudinal polarization.    

Let us count the physical degrees of freedom in 4 dimensions. Since the polarization tensor $\xi_3$ is antisymmetric, the total number of polarization before taking into account the physical state condition is  $4+4\times 4+4=24$. The physical state condition (\ref{constrain}) imposes $11$ conditions, so we have $24-11=13$ physical degree of freedom.  In 4 dimension, a massive spin $J$ particle has $2J+1$ physical degrees of freedom.  From the calculations in a previous work \cite{AGT3}, we know the $n=1$ string modes include two $J=0$ particles, one of which decays exclusively to two gluons with $(++)$ helicity and the other to two gluons with $(--)$ helicity. Therefore, 
the 13 degrees of freedom correspond naturally to that of one $J=2$, two $J=1$, and two $J=0$ string resonance modes.

Using the state-operator correspondence (\ref{stateoperator}), we can easily write the vertex operator for the $n=1$ string mode (\ref{stringstate}) in the $(-1)$ picture  as follows:
\begin{eqnarray}
	\mathcal{O}(z) = [(\xi_1)_\mu \partial\psi^\mu +\frac{i}{(2\alpha^\prime)^{\frac{1}{2}}}(\xi_2)_{\mu\nu}\partial X^\mu \psi^\nu
	+(\xi_3)_{\mu\nu\rho} \psi^\mu\psi^\nu\psi^\rho] T^a e^{-\phi(z)}e^{ik\cdot X(z)}
\end{eqnarray}

\end{document}